\documentstyle[aps,twocolumn,floats,prl]{revtex}

\input epsf

\newcommand{\Nnu}{N}
\newcommand{\eV}{{\rm\,eV}}

\newcommand{\ihmpc}{h{\rm\,Mpc}^{-1}}

\def\kmin{k_{\rm min}}
\def\kmax{k_{\rm max}}
\def\F{{\bf F}}

\begin{document}
\twocolumn[\hsize\textwidth\columnwidth\hsize\csname
@twocolumnfalse\endcsname

\title{Weighing Neutrinos with Galaxy Surveys}
\author{Wayne Hu,
Daniel J. Eisenstein, 
\& Max Tegmark
}
\address{Institute for Advanced Study, School of Natural Sciences,
Princeton, NJ 08540 }

\maketitle 
\begin{abstract}We show that galaxy redshift
surveys sensitively probe the neutrino 
mass, with eV mass neutrinos suppressing power by a factor of
two.  The Sloan Digital Sky Survey can potentially
detect $N$ nearly degenerate massive neutrino species with mass $m_\nu \agt
0.65 (\Omega_m h^2/0.1 \Nnu)^{0.8} \eV$ at better than $2\sigma$ once 
microwave background experiments measure two other cosmological parameters.
Significant overlap exists between this region and that implied by the
LSND experiment, and even $m_\nu \sim 0.01-0.1\eV$, as implied by the
atmospheric anomaly, can affect cosmological measurements.
\end{abstract}
]


Current neutrino experiments reveal anomalies resolvable by nonzero 
neutrino masses and flavor oscillations.
The LSND direct detection experiment suggests $\nu_\mu$ to $\nu_e$ oscillations
with $\delta m^2_{\mu e} \agt 0.2\eV^2$ \cite{LSND}.  
The deficit of $\mu$ neutrinos in atmospheric showers indicates mixing
between $\nu_\mu$ and another species with 
$\delta m^2_{\mu i} \sim 10^{-3}-10^{-2}\eV^2$ \cite{Atmos}.  
Finally, the solar neutrino deficit requires $\delta
m^2_{e i} \sim 10^{-5}\eV^2$ (see e.g. \cite{Solar} for recent
assessments).  These results are consistent with one to three
weakly interacting neutrinos in the eV mass range \cite{Sterile}.

Cosmological measurements provide an independent, albeit indirect \cite{BBN},
means of determining neutrino masses in the above-mentioned range. 
Massive neutrinos
would produce a strong suppression in the clustering of galaxies, with even
a 10\% neutrino contribution making a 100\% difference in the power 
\cite{Hu97}.  
Detecting this suppression would measure the
absolute mass of the neutrinos, in contrast to the mass splittings measured by
the oscillation effects described above.

While the general effect is well known, most work to date has 
focused on a combined neutrino mass around $5\eV$, as this
is the minimum needed to affect cosmology at the current
observational sensitivities \cite{DLA,Pri95}.
There are three reasons why this situation
is likely to change soon.  First, evidence continues to mount
that we live in a low-density universe (e.g. \cite{LowOmega}).
Since the cosmological effects depend on the density {\it fraction} 
supplied by neutrinos, our sensitivity to the neutrino mass increases
roughly in inverse proportion to the density parameter.
Second, the cosmic microwave background (CMB) 
experiments currently under development
should establish a cosmological framework (e.g. \cite{Nature})
that is as secure as the standard model of
particle physics.  
The parameters left unspecified
by the model may then be measured with confidence.  
Finally, upcoming high precision galaxy 
surveys such as the
Sloan Digital Sky Survey (SDSS) \cite{SDSS} should be able to measure the
total power on the relevant scales to $\sim$1\% accuracy.  The combination
of these developments implies that galaxy surveys will soon 
provide either an interesting constraint on or a detection of
the mass of the neutrinos. 

Although the effects of massive neutrinos are very large, variations
in other cosmological parameters may mimic the signal.  Therefore,
to qualify as a true detection, all other
aspects of the cosmology that similarly affect the power spectrum must
be previously or simultaneously determined.

In this {\it Letter}, we evaluate the ability of galaxy surveys
to distinguish between these possibilities and
thereby measure the mass of the neutrinos. 
We establish the physical basis
of the measurement,				 
evaluate 			  	 
the uncertainties caused by our ignorance
of other aspects of cosmology, and   
present the
$2\sigma$ detection threshold in mass for SDSS.
These results depend on two assumptions: that CMB observations
will confirm that
structure forms through the gravitational instability of cold dark matter,
and that the galaxy bias is linear, i.e., the galaxy power
spectrum is proportional
to the underlying mass power spectrum.  
The second assumption is relaxed in the concluding remarks.


{\it Neutrino Signature.\ ---} 
In a universe with the standard thermal history \cite{Kol90}, the
temperature of the background neutrinos is $(4/11)^{1/3}$ that of the CMB.  
This implies $\Omega_\nu h^2 \approx \Nnu
m_\nu/94\eV$ 
for $\Nnu$ massive neutrino species of nearly identical mass $m_\nu$.
Here and below, $\Omega_i$ is the fraction of the critical density
contributed by the $i$th matter species ($\nu=$ neutrinos, $b=$
baryons, $m=$ all matter species) and $H_0 = 100h$ km s$^{-1}$
Mpc$^{-1}$.  We assume this thermal history and 
a power-law spectrum of initial adiabatic density fluctuations with $P(k)
\propto k^n$ throughout.

\begin{figure}[t]
\begin{center}
\leavevmode
\epsfxsize=3.4truein \epsfbox{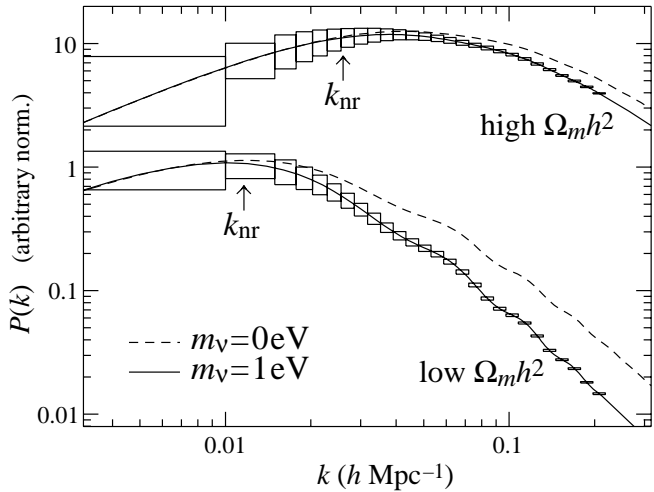}
\end{center}
\caption{Effect of a 1 eV neutrino on the BRG power spectrum
compared with expected precision of the SDSS BRG survey 
(1$\sigma$ error boxes, assuming $\sigma_8=2$ for
the BRGs). Upper curves:
an 
$\Omega_m=1.0$, $h=0.5$,
$\Omega_b h^2=0.0125$, $n=1$ 
model
with and without a 1 eV neutrino mass.
Lower curves: the same but for an 
$\Omega_m=0.2$, $h=0.65$ model.  
}
\label{fig:neutrino_1}
\end{figure}

This initial power spectrum is
processed by the gravitational instability of the fluctuations. 
The large momentum of cosmological eV mass neutrinos
prevents them from clustering with the cold components on scales smaller
than the neutrinos can move in a Hubble time.  
The growth of the fluctuations is therefore suppressed on 
all scales below the horizon when the neutrinos become nonrelativistic \cite{Hu97}
\begin{equation}
k_{\rm nr} \approx 0.026 \left( {m_\nu \over 1\eV} \right)^{1/2} 
	\Omega_m^{1/2} h {\rm\,Mpc}^{-1}.
\label{eqn:kfs}
\end{equation}
The small-scale suppression is given by 
\begin{equation}
\left( {\Delta P\over P }\right) \approx -8{\Omega_\nu\over\Omega_m}\approx-0.8
\left( { m_\nu \over 1\eV} \right) \left( {0.1\Nnu \over\Omega_m h^2}\right)\,.
\label{eqn:delP}
\end{equation}
Galaxy surveys such as the SDSS Bright Red Galaxy (BRG) survey 
(assumed to be volume-limited to $1h^{-1}{\rm\,Gpc}$ \cite{SDSS}) 
should measure the power between $0.1-0.2h{\rm\,Mpc}^{-1}$ to $\sim$$1\%$.
A more detailed analysis shows that only masses below
\begin{equation}
m_{\rm min} \approx 0.02 (\Omega_m h^2/0.1\Nnu)\eV
\label{eqn:mmin}
\end{equation}
make less than a $2\sigma$ change in the power spectrum measured by the
BRG survey.

As an example, we plot in Fig.\ \ref{fig:neutrino_1} the power
spectrum with and without a single 1 eV massive neutrino species for an
$\Omega_m=0.2$, $h=0.65$ model (lower curves) and an $\Omega_m=1.0$,
$h=0.5$ model (upper curves).  The
expected $1 \sigma$ error boxes from the BRG survey 
shows that the two models are clearly distinguishable.
For comparison, the
difference between these models in the CMB power spectrum at degree
angular scales is roughly 3\% (1\%) and never exceeds 5\% (4\%) 
for multipoles $\ell < 2000$ for the open variant of the low
(high) $\Omega_0 h^2$ cases (c.f.~\cite{BoltMDM}).


{\it Parameter Degeneracies.\ ---} 
Although the suppression of power caused by massive neutrinos is large,
we must consider whether other cosmological effects can
mimic this signal.  The suppression 
begins at $k_{\rm nr}$ [Eq.~(\ref{eqn:kfs})]
and approaches the constant factor of
Eq.~(\ref{eqn:delP}) at smaller scales.  Many cosmological effects
can produce the gross effect of a change in the ratio of large to small
scale power; we must rely on the detailed differences between these
mechanisms in order to distinguish one from another.

\begin{figure}[bt]
\begin{center}
\leavevmode
\epsfxsize=3.4truein \epsfbox{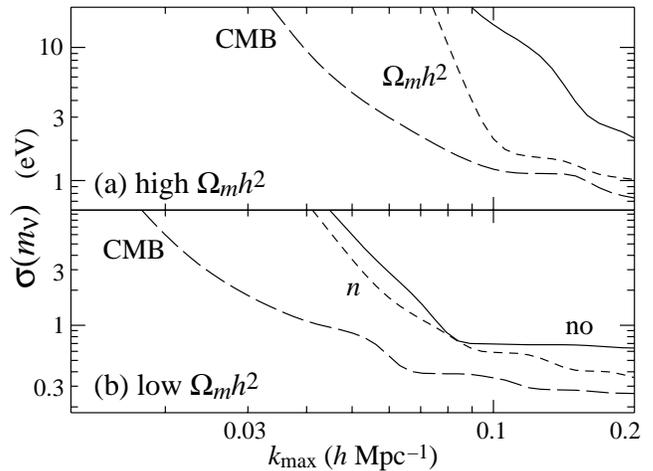}
\end{center}
\caption{Standard deviation $\sigma({m_\nu})$ as a function of the
upper cutoff $\kmax$ for several different choices of prior
cosmological constraints.  Models are the same as in
Fig.~\protect\ref{fig:neutrino_1} and have $m_\nu=1\eV$.  
({\it a}) High $\Omega_m h^2$:  no priors, solid line; single prior of
$\sigma(\Omega_m h^2)=0.04$, dashed line; full CMB prior (see text),
long-dashed line.  
({\it b}) Low $\Omega_m h^2$:  as ({\it a}), save
that the single prior is $\sigma(n)=0.06$, dashed line.  
}
\label{fig:neutrino_2}
\end{figure}

We consider variants of the adiabatic cold dark matter model.
The power spectrum is then 
a function of the normalization $A$, tilt $n$, $h$, $\Omega_m h^2$,
$\Omega_b h^2$ and $\Omega_\nu h^2$.  The spatial curvature, cosmological constant,
and the linear bias parameter are implicitly included through the
normalization \cite{Hu97}.  
We estimate the accuracy with which these parameters can be jointly
measured from the SDSS BRG survey using the technique described in
\cite{Teg97}. Here the $6\times 6$ covariance matrix of the 6 
parameter estimates is approximated by the inverse of the so-called 
Fisher information matrix $\F$.  Its elements ${\bf F}_{ij}$ are obtained by
integrating $(\partial_i \ln P)\, (\partial_j \ln P)$ over the range
$\kmin\le k\le\kmax$ discussed below, 
weighted by a function that incorporates the relevant aspects of the survey
geometry and sampling density.  Here the derivatives are 
with respect to the $i^{th}$ and $j^{th}$ parameters, evaluated at a fiducial model.

If a small variation in one parameter can be mimicked by joint
variations in other parameters, then one of these functions $\partial_i \ln P$ 
can be approximated by a linear combination of the others.
This situation is referred to as a {\it parameter degeneracy}, since  
it makes $\F$ nearly singular and leads to extremely poor determinations
(large variance $\F^{-1}_{ii}$) for the parameters involved.

Clearly, the ability to estimate parameters comes only from scales
on which one has both precise measurements and reliable theoretical
predictions.
On large scales, linear perturbation theory is accurate, but the survey
volume (about $1h^{-3}$~Gpc$^3$ for the BRG survey) is limited; hence
$\kmin\approx 0.005\ihmpc$.  On small scales, linear theory fails
near $k\approx0.2\ihmpc$.  While we expect that detailed data analysis
will push $\kmax$ to slightly smaller scales by including mild
corrections to linear theory, we simply use the linear
power spectrum for this work and adopt $\kmax\approx 0.2\ihmpc$.
We shall see that if $\kmin \alt
k_{\rm nr} \alt \kmax$, then the unique signature of massive neutrinos can
be identified and $m_\nu$ measured. 

The solid lines in Fig.\ \ref{fig:neutrino_2} show the standard
deviation of a measurement of 
$m_\nu$ (or, equivalently, $\Omega_\nu h^2$) as a function
of $\kmax$ if all relevant cosmological parameters are determined
simultaneously from the SDSS BRG data set.  Consider first the
low $\Omega_m h^2$ case (bottom panel).  The standard deviation
drops rapidly near $\kmax=0.05\ihmpc$, well below the scale at
which the neutrinos begin to affect the power spectrum 
(c.f.\ Fig.~\ref{fig:neutrino_1}).  
If we use only information from $k\alt0.05\ihmpc$, we 
find that the neutrino signal can be accurately duplicated by
variations in other parameters.  For example, a change in normalization
and tilt would be indistinguishable within the BRG survey error bars.
When considering smaller scales, more subtle combinations still
exist; these near degeneracies reduce the parameter
sensitivity more than 100-fold.  
A similar situation occurs in the high $\Omega_m h^2$ case (top panel)
but at a smaller scale [Eq.\ (\ref{eqn:kfs})].

If we possess external information on the other cosmological parameters,
the situation improves dramatically because parameters may no longer be
shifted arbitrarily so as to mimic the neutrino signal.
Indeed, upcoming CMB anisotropy experiments 
should yield precise measurements of cosmological parameters critical
to this situation.
We therefore show the effect (long-dashed curve) of
placing CMB constraints on the cosmological parameters:
$\sigma({\ln A})=0.40$, $\sigma(n)=0.06$, $\sigma({\Omega_m h^2})=0.04$,
$\sigma({\Omega_b h^2}) = 0.1\Omega_bh^2$, and $\sigma(h)=0.1$, where
$\sigma(i)$ denotes the standard deviation of $i$.
We view these constraints as quite conservative, since they 
are weaker than those predicted for the 
MAP satellite \cite{Bon97,Hat97} and ignore the
tight correlation between the marginalized error bars \cite{Complementarity}.

Which one prior is most important depends upon the fiducial model.
For low $\Omega_m h^2$
models with small neutrino fractions, one cannot accurately probe
the scales on which the neutrino suppression is small since $k_{\rm min} \agt
k_{\rm nr}$.
This enables the tilt $n$ to produce much of the desired effect.  
We show the error bars resulting from including only the tilt
prior in Fig.\ \ref{fig:neutrino_2} (short-dash line).  While this is the most
important prior at $\kmax=0.2\ihmpc$, the others combined have a non-negligible 
effect.

For the high $\Omega_m h^2$ case, one has precise measurements
around $k_{\rm nr}$, so that the onset of neutrino effects can be
distinguished from tilt (c.f.\ Fig.\ \ref{fig:neutrino_1}).
However, altering $\Omega_m h^2$ or $h$ causes $P(k)$ to
slide horizontally (leaving the largest scales unchanged); as
one can see in Fig.\ \ref{fig:neutrino_1}, this is roughly degenerate with the neutrino
effect.  The 
$\Omega_m h^2$ prior is most important in this case; we
show this situation in Fig.\ \ref{fig:neutrino_2} (upper panel, short-dash line).

We also test how $\sigma(m_\nu)$ increases as we double each prior in turn. 
The results change by more than a few percent only for
$\Omega_m h^2$ (20\%) and $n$ (10\%) in the high $\Omega_m h^2$ model
and for $n$ (40\%) in the low $\Omega_m h^2$ model.


\begin{figure}[t]
\begin{center}
\leavevmode
\epsfxsize=3.5truein \epsfbox{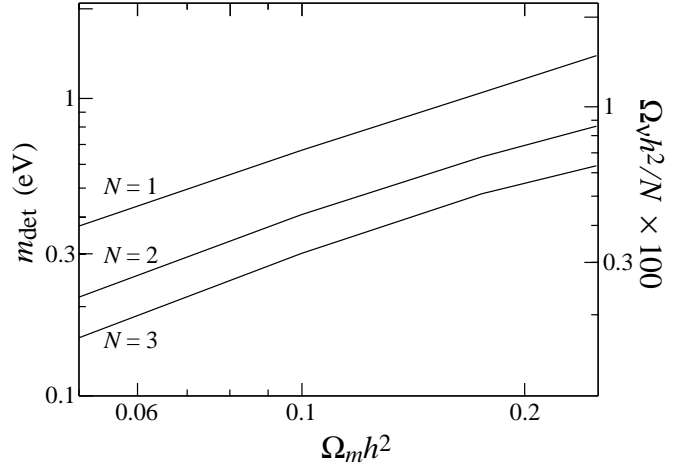}
\end{center}
\caption{The $2\sigma$ detection threshold for $m_\nu$ from the SDSS
BRG survey as a function of the matter density $\Omega_m h^2$ for the
number of degenerate mass neutrinos $\Nnu=1$--3.  
We have used $h=0.5$, $\Omega_bh^2=0.0125$, $n=1$, and $\kmax=0.2\ihmpc$;
variations on these produce only mild shifts.
}
\label{fig:neutrino_3}
\end{figure}

{\it Results.\ ---} Given the confusion with variations in other cosmological parameters,
what is the minimum detectable neutrino mass $m_{\rm det}$?
In Fig.~\ref{fig:neutrino_3}, we show the $2\sigma$ detection
threshold [i.e.\ $m_\nu = 2\sigma(m_\nu)$]
assuming the CMB priors given above, $\kmax=0.2\ihmpc$, 
and a family of fiducial models with $\Omega_b h^2=0.0125$,
$h=0.5$, and $n=1$.  The choice of a 
fiducial model does not amount to fixing cosmological parameters;
all parameters are determined by the galaxy data or by the
prior constraints.  

With these choices, SDSS can detect the neutrino(s) if
\begin{equation}\label{eqn:mlimit}
m_\nu \agt m_{\rm det} \approx 0.65 \left(\Omega_mh^2\over 0.1\Nnu\right)^{0.8}\eV.
\end{equation}
If the exponent here were unity, it would correspond to 
a fixed fractional suppression of power [Eq.~(\ref{eqn:delP})].
In practice, one does slightly better at larger $\Omega_mh^2/\Nnu$ because 
$k_{\rm nr}$ [Eq.\ (\ref{eqn:kfs})] is larger and thus better resolved.

This result is fairly insensitive to changes in the 
fiducial model or survey parameters.
Choosing $h=0.8$ increases $m_{\rm det}$ by 15\% at low $\Omega_mh^2$;
doubling the baryon density does the opposite.  Neither matters at
high $\Omega_mh^2$.  Altering $n$ or $A$ affects the answer very little. 
Reverting from the deeper
BRG survey to the main SDSS North survey \cite{SDSS}
increases $m_{\rm det}$ by less than 25\%.  

As for the assumptions implicit in Fig.~\ref{fig:neutrino_3},
only the prior constraints on the tilt
$n$ in the low $\Omega_0 h^2$ regime and $\Omega_0 h^2$ itself in the
high regime are essential.  We have taken conservative priors from the CMB 
here and save a full joint analysis for future work \cite{Complementarity}. 
Decreasing $\kmax$ to $0.13\ihmpc$ increases $m_{\rm det}$ 
by $\sim$$40\%$ at large $\Omega_mh^2$ but makes little difference at
small $\Omega_mh^2$ (c.f.\ Fig.\ \ref{fig:neutrino_2}).

Quasi-linear evolution near $\kmax$ presents a complication to
the analysis, but so long as the power spectrum can be
calculated as a function of cosmological parameters through simulations
or analytic approximations \cite{HKLM}, this need not necessarily
degrade the parameter estimation.  However, to the extent that
evolution washes out features in the power spectrum, degeneracies may
appear that are not present in linear calculation.  
Hence, this issue merits
further investigation, although we view our choice of $\kmax$ as
conservative.  Fortunately, as shown in Fig.\ \ref{fig:neutrino_2},
prior information from the CMB assists in making the results robust
against changes in $\kmax$.

Our fundamental assumption is that the power spectrum of the galaxies
is proportional to that of the underlying mass, i.e., that
the galaxy bias is linear.  This assumption is well-motivated 
in the linear regime \cite{LinearBias}.  
Bias that develops a scale dependence as fluctuations become non-linear introduces 
only moderate uncertainties because it does not accurately mimic the neutrino
signature.  Adopting the prescription of \cite{Man97} [their eq. (20)]
and
marginalizing over the scale-dependence parameter $b_2$ 
degrades the result in the models of Fig.~\ref{fig:neutrino_2} 
by a factor of 2.5 implying a similar effect on 
the detection threshold of Eq.~(\ref{eqn:mlimit}).
The degradation stems not from the presence of
non-linear bias but from our ignorance of its amplitude.  The latter can be constrained
by the scale dependence of redshift space distortions and 
the {\it relative} bias of different galaxy populations.  
Peacock \cite{Pea97} determines $\sigma(b_2)=0.01$ for the relative bias of
IRAS and optical galaxies; bounding scale-dependent bias at this level
would restore the limits of Eq.~(\ref{eqn:mlimit}).

In summary, although galaxy surveys can measure power to $\sim$$1\%$, 
isolating the mass of the neutrino at 2$\sigma$ requires $\sim$$50\%$
power variations.  In principle, this means that a 50-fold improvement
of $m_{\rm det}$ 
to $m_{\rm min}$ [Eq.~(\ref{eqn:mmin})] 
would be available if other cosmological parameters were known
perfectly.  While this is unrealistic, a measurement with a precision
of $\sigma(\ln A) \sim 0.1$ and $\sigma(n)\sim0.03$ would yield a
factor of two improvement in $m_{\rm det}$ and is potentially within
reach of planned experiments.

For $m_\nu$ between $m_{\rm min}$ and 
$m_{\rm det}$, the effects on the power spectrum are significant yet cannot
be robustly attributed to massive neutrinos.  
As this brackets the mass range implied by atmospheric neutrinos,
the possibility of a light massive neutrino species must
be considered when measuring other cosmological parameters.

Massive neutrinos present an example where the galaxy power spectrum
provides cosmological information on fundamental physics not available in CMB
measurements, but where CMB measurements are are nonetheless needed
for an unambiguous detection. 
This illustrates the complementary nature of galaxy surveys and CMB anisotropies.

\smallskip
We thank J.N.\ Bahcall, E.\ Lisi, and N.\ Hata for useful discussions
and acknowledge use of CMBfast \cite{CMBfast}.  WH and DJE were
supported by NSF PHY-9513835, WH by a Sloan Fellowship and
the WM Keck Foundation, DJE by the Frank and Peggy Taplin Membership, 
and MT by NASA through Hubble Fellowship
HF-01084.01-96A, from STScI operated by AURA, Inc. 
under NASA contract NAS5-26555.



\begin{thebibliography}{99}
\frenchspacing

\bibitem{LSND} C. Athanassopoulos et al., 
	Phys. Rev. Lett. {\bf 75}, 2650 (1995);
	ibid, Phys Rev. C {\bf 54}, 2685 (1996)
\bibitem{Atmos} 
	Gaisser, T. K., Halzen, F., and Stanev, T., Phys. Rep. {\bf 258}, 173 (1995)
\bibitem{Solar} J. N. Bahcall, Astrophys. J. {\bf 467}, 475 (1996);
	N. Hata and P. Langacker, Phys. Rev. D {\bf 56}, 6107 (1997)
\bibitem{Sterile}
	A sterile species may be required. See
	K. S. Babu, R. K. Schaefer, and Q. Shafi,  
	Phys. Rev. D {\bf 53}, 606 (1996);
	G. L. Fogli, E. Lisi, D. Montanino, and G. Scioscia, 
	Phys. Rev. D {\bf 56}, 4365 (1997)
\bibitem{BBN}
	Any particle of comparable mass has similar effects.
\bibitem{Hu97} W. Hu and D. J. Eisenstein, Astrophys. J. {\bf 498}, 497 (1998);
 	D. J. Eisenstein and W. Hu, astro-ph/9710252
\bibitem{DLA} H. J. Mo and J. Miralda-Escud\'e, 
	Astrophys. J. {\bf 430}, L25 (1994);
	G. Kauffmann and S. Charlot, {\it ibid}, {\bf 430}, L97 (1994);
	A. Klypin, S. Borgani, J. A. Holtzman, and J. R. Primack, {\it ibid},
	{\bf 441}, 1 (1995)
\bibitem{Pri95}
	J. R. Primack, J. Holtzman, A. Klypin, and D. O. Caldwell, 
	Phys. Rev. Lett. {\bf 74}, 2160 (1995) 	
\bibitem{LowOmega} N. A. Bahcall, X. Fan, and R. Cen, Astrophys. J. 
	{\bf 485}, 53 (1997); M. Bartelmann et al., astro-ph/9707167.
\bibitem{Nature} W. Hu, N. Sugiyama, J. Silk, Nature {\bf 386}, 37 (1997)
\bibitem{SDSS} SDSS: http://www.astro.princeton.edu/BBOOK 
\bibitem{Kol90} E. Kolb and M. Turner, 
	{\it The Early Universe} (Addison-Wesley, New York, 1990) 
\bibitem{BoltMDM}
	C.-P. Ma and E. Bertschinger, Astrophys. J. {\bf 455}, 7 (1995);
	S. Dodelson, E. Gates, and A. Stebbins, 
	{\it ibid}, {\bf 467}, 10 (1996)
\bibitem{Teg97} M. Tegmark, Phys. Rev. Lett. {\bf 79}, 3806 (1997)
\bibitem{Bon97} J. R. Bond, G. Efstathiou, and M. Tegmark, 
	Mon. Not. Roy. Astron. Soc., {\bf 291}, 33 (1997);
        M. Zaldarriaga, D. N. Spergel, and U. Seljak, Astrophys. J., {\bf 488} 1 (1997);
	G. Jungman, M. Kamionkowski, A. Kosowsky, and D. N. Spergel,
	Phys. Rev. D {\bf 54}, 1332 (1996)
\bibitem{Hat97} $\sigma(\ln A)$ employs
         $\Omega_m^{0.6}/b$ from SDSS; see
	 S. Hatton \& S. Cole, Mon. Not. Roy. Astron. Soc. {\bf 296}, 10 (1998)
\bibitem{Complementarity} Correlations improve the limits by $\sim 2$; 
	see D.J. Eisenstein, W. Hu, \& M. Tegmark (in preparation)	
\bibitem{HKLM} A. J. S. Hamilton, P. Kumar, E. Lu, and A. Matthews, 
	Astrophys. J. {\bf 374}, 1 (1991) 
\bibitem{LinearBias} P. Coles, Mon. Not. Roy. Astr. Soc. {\bf 262}, 1065 (1993);
	R.J. Scherrer, D.H. Weinberg, preprint, astro-ph/9712192 
\bibitem{Man97} 
	R.G. Mann, J.A. Peacock, A.F. Heavans, Mon. Not. Roy. Astr. Soc.
	{\bf 293}, 209 (1998);
	we express their prescription as 
	$1+\Delta_{\rm gal}^2 = [1+(b_1^2/b_2)\Delta_{\rm mass}^2]^{b_2}$
\bibitem{Pea97} J.A. Peacock, Mon. Not. Roy. Astr. Soc. {\bf 284} 885 (1997) 
\bibitem{CMBfast}
	U. Seljak and M. Zaldarriaga, Astrophys. J. {\bf 469}, 437 (1996)

\nonfrenchspacing
\end{thebibliography}
\end{document}